\documentclass[12pt]{article}
\usepackage{amsmath,amssymb,epsf,epsfig,amsthm,bm,graphicx}
\usepackage[dvips]{color}

\begin{document}

\title{ \flushright{\small CECS-PH-07/03} \\
\center{The Universe as a topological defect}}
\author{\textbf{Andr\'{e}s Anabal\'{o}n$^{1,2}$}\thanks{%
anabalon-at-cecs-dot-cl} \textbf{\ \ \ \ Steven Willison$^{1}$}\thanks{%
steve-at-cecs-dot-cl} \and \textbf{Jorge Zanelli$^{1}$}\thanks{%
jz-at-cecs-dot-cl} \\
\textbf{\textit{$^{1}$}}{\small Centro de Estudios Cient\'{\i}ficos (CECS)
Casilla 1469 Valdivia, Chile}\\
{\small \textit{${}$}}\textbf{\textit{$^{2}$}} {\small Departmento de F\'{\i}%
sica, Universidad de Concepci\'{o}n Casilla 160-C, Concepci\'{o}n, Chile }\\
}
\maketitle

\begin{abstract}
Four-dimensional Einstein's General Relativity is shown to arise from a
gauge theory for the conformal group, SO(4,2). The theory is constructed
from a topological dimensional reduction of the six-dimensional Euler
density integrated over a manifold with a four-dimensional topological
defect. The resulting action is a four-dimensional theory defined by a
gauged Wess-Zumino-Witten term. An ansatz is found which reduces the full
set of field equations to those of Einstein's General Relativity. When the
same ansatz is replaced in the action, the gauged WZW term reduces to the
Einstein-Hilbert action. Furthermore, the unique coupling constant in the
action can be shown to take integer values if the fields are allowed to be
analytically continued to complex values.
\end{abstract}


\section{Introduction}


Besides its observational success in the solar system, in measurements of
the binary pulsar, and in the early universe through primordial
nucleosynthesis, Einstein's general relativity (\textbf{GR}) has a beautiful
mathematical formulation. One of the appealing mathematical features is its
connection with a topological invariant in two dimensions. The well known
relation of the Einstein-Hilbert Lagrangian and the Euler characteristic can
be summarized as follows:
\begin{equation}
S_{EH}=\frac{c^{3}}{16\pi G}\zeta_{4}(M),\; \chi _{2}(M)=\frac{1}{4\pi}
\zeta _{2}(M),\qquad \zeta _{D}(M)=\int_{M}R\sqrt{\left\vert g\right\vert}
d^{D}x.
\end{equation}
This fact, sometimes referred to as the dimensional continuation of the
Euler density, has a straight-forward generalization to higher dimensions,
giving rise to the Lovelock series \cite{Lovelock:1971yv, Zumino:1985dp}.
This series in dimension $D$ contains $\left[ \frac{D+1}{2}\right] $ terms,
where $[\cdots ]$ denotes the integer part. The terms are the dimensionally
continued Euler densities of all dimensions below $D$, and the cosmological
constant term.

Although the dimensional continuation process gives a well defined
prescription to obtain the most general, ghost-free\footnote{%
For perturbations around flat space.}, gravitational Lagrangian \cite%
{Zwiebach:1985uq}, its Kaluza-Klein (\textbf{KK}) reduction to four
dimensions gives standard GR with an arbitrary cosmological constant and
with additional constraints that force, for instance, the four dimensional
Euler density to vanish \cite{Mueller-Hoissen:1985ij,Giribet:2006ec}. This
is a generic feature of the dimensional reduction of theories that contain
higher powers of curvature. It is commonly believed that higher curvature
corrections to the Einstein-Hilbert action produce small deviations from GR,
but this is actually not true: the field equations, obtained from the
variation of the reduced action with respect to the four-dimensional
scalars, produce constraints additional to the Einstein equations which rule
out many solutions of GR, including the gravitational field of a spherically
symmetric source \cite{EHTZ}.

This problem is analogous to the one encountered in the gauge theory sector
in standard KK reductions to four dimensions starting from the
Einstein-Hilbert action in $D>4$, where the Yang-Mills density must
necessarily vanish in backgrounds with constant scalars. Thus, although the
behavior of theories obtained by the KK reduction of Lovelock Lagrangians
could be reasonable at the galactic scale or at the beginning of our
Universe, at the scale of our solar system their departure from the GR
behavior is not experimentally acceptable. On the other hand, there is the
largely unsolved problem of the non-renormalizability, in the power counting
sense \cite{Gomis:1995jp}, of the gravitational interaction. Although pure
gravity has a finite one-loop $S$ matrix \cite{'tHooft:1974bx}, until now
all matter couplings --except supergravity \cite{Grisaru:1976ua}--, destroy
this one-loop behavior. At two loops, pure gravity diverges \cite%
{Goroff:1985th}, and at three loops also supergravity contains divergences
\cite{Deser:1978br}, although the coefficient in front of the divergence has
not been computed until now \cite{Bern:2002kj}. One is left with an
uncomfortable scenario, in which there is no field theory formulation to
compute a simple graviton scattering in a consistent way. These facts
motivate the search for new theories that include Einstein's field equations
in some way, but that also contain other dynamical sectors, such that other
phenomena can be explained within these theories.

A useful guide can be found in the three dimensional case which, in the
first order formalism, can be seen as a gauge theory, where the vielbein $e$
and the spin connection $\omega$ are part of a single connection \cite%
{Achucarro:1987vz}. This Chern-Simons (\textbf{CS}) theory for gravity
contains a larger set of field configurations than metric GR. Indeed, by a
gauge transformation any of the components of a flat connection can always
be set equal to zero in an open neighborhood. Thus, a generic field
configuration of CS gravity does not have a metric interpretation.
Projection of the gauge theory to the sector where the vielbein is
invertible and the connection is torsion-free, allows one to recover the
usual metric theory of gravity.

Three-dimensional CS theory is renormalizable, as follows from the fact that
the unique dimensionless coupling constant can only take integer values (in
fact, it is finite at the quantum level) \cite{Witten:1988hc}, \cite%
{Deser:1989xu}. Renormalization of three-dimensional gravity can then be
proven by embedding the theory in a gauge theory with principal bundle
structure, in accordance with the fact that all known physical interactions
which make sense quantum mechanically are explained by gauge theories. Thus,
an embedding of four dimensional GR in a gauge theory where $e$ and $\omega$
are parts of a single connection, is a welcome feature.

The theoretical motivation is quite natural. Instead of considering the
dimensional continuation of the two dimensional Euler density, the four
dimensional Lagrangian will be given by a topologically induced dimensional
reduction of the six dimensional Euler density. The dimensional reduction
mechanism occurs due to the introduction of a four-dimensional topological
defect in the six dimensional manifold where the Euler density is
integrated. This approach was already studied in \cite%
{Bonelli:2000tz,Boyarsky:2002gi}. Those authors, however, restrict the
connection in the action such that the only degrees of freedom left at the
defect are the components which correspond to the four-dimensional $e$ and $%
\omega$, obtaining in this way, just the usual Einstein-Hilbert plus
cosmological constant action.

Here, instead, no restrictions are imposed in the reduction process and the
non triviality of the bundle is always assumed. This gives rise to a
four-dimensional theory with a lagrangian that is gauge invariant under the
conformal group $SO(4,2)$. This symmetry is broken down to $SO(3,1)$ by the
presence of the defect. The theory is defined by the metric-independent
sector of the gauged Wess-Zumino-Witten (gWZW) action. The kinetic term $%
Tr(D_{\mu} g D^{\mu}g^{-1})$ --where $D_{\mu }g=\partial _{\mu}g +\left[
\mathcal{A}_{\mu}, g \right]$, $Tr$ is the bilinear invariant of the Lie
group, and $\mathcal{A}_{\mu}$ is the Lie algebra valued connection--, never
arises in our construction \cite{Anabalon:2006fj}. The resulting action
resembles in many ways its three-dimensional, quantum mechanically finite
sibling: in both cases $e$ and $\omega $ are part of a single $SO(m,n)$
connection $\mathcal{A}$; both theories admit a vacuum configuration $%
e=\omega =0$, in which the space-time causal structure completely
disappears; both have a quantized dimensionless ``coupling" constant in
front of the action. The discreteness of this constant makes any continuous
process of renormalization impossible, hinting that the beta function must
be zero.

In Section 2, the mechanism of dimensional reduction is discussed. For the
sake of simplicity, the discussion is presented first analyzing the
four-dimensional Euler density integrated on a four-dimensional spacetime
with a two-dimensional defect. The extension of results to reduce from six
to four dimensions together with the field equations, is stated. In Sect. 3,
the on shell configuration that reproduces Einstein's gravity is discussed.
The conditions under which the coupling constant takes integer values are
discussed in Sect. 4, and Sect. 5 contains the discussion and conclusions.


\section{Topologically induced dimensional reduction}


Observing that four dimensional gravity is the dimensional continuation of
the two-dimensional Euler density, the natural object to dimensionally
reduce is the six-dimensional Euler density\footnote{%
In this work the exterior product between forms is omitted, i.e $F\wedge
F\equiv FF$. Since pullback and exterior derivatives commute, they are
usually omitted in physics literature, and we follow that convention. For
more conventions see the appendix.},
\begin{equation}
\chi (M)=\frac{1}{48\pi ^{3}}\int_{M^{6}}\left\langle \mathcal{FFF}%
\right\rangle =\frac{1}{48\pi ^{3}}\frac{1}{2^{3}}\int_{M^{6}}\varepsilon
_{ABCDEF}F^{AB}F^{CD}F^{EF},  \label{Euler}
\end{equation}%
where the indices $A,B,...$ go from $0$ to $5$, $\mathcal{F}=\frac{1}{2}%
J_{AB}F^{AB}=d\mathcal{A}+\mathcal{AA}$ is the pseudo-Riemannian curvature
of the six-dimensional manifold\footnote{%
We call it $\mathcal{F}$ so as not to confuse it with its four dimensional
analog $\mathcal{R}$.}. Depending on the signature of the six dimensional
metric, the generators $J_{AB}$ can be assumed to span any of the algebras $%
so(6),$ $so(5,1)$, $so(4,2)$ or $so(3,3)$. The symmetric trace $\left\langle
...\right\rangle $ is the Levi-Civitta invariant tensor of these groups, $%
\left\langle J_{AB}J_{CD}J_{EF}\right\rangle =\varepsilon _{ABCDEF}\,$ and $%
\partial M^{6}=\varnothing $. As will be shown, a dimensional reduction
occurs if a four-dimensional sub-manifold is removed from $M^{6}$, producing
a topological defect. However, in order to be able to use the standard
exterior calculus (e. g., Stokes Theorem), and pass from the six-dimensional
integral to a four-dimensional one, a limiting process is needed. Here the
topological defect will be created by removing a six-dimensional cylinder $%
M^{4}\times D^{2}$, and then taking the limit in which the radius of the two
dimensional disc $D^{2}$ shrinks to zero. This is known as a regularization
process to remove a sub-manifold of codimension two.
\begin{figure}[t]
\centering
\parbox{.45\textwidth}{\includegraphics[width=.49\textwidth]{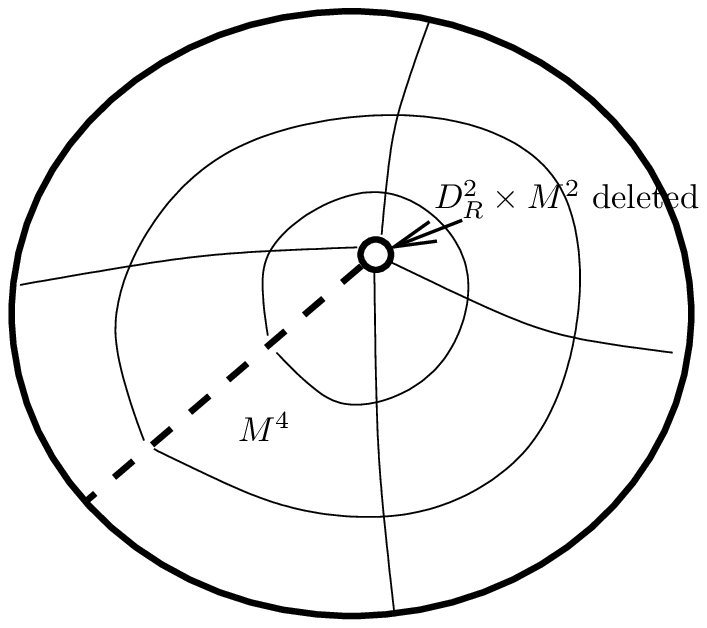}
     \caption{ {\small A two-dimensional defect in a four-dimensional
     manifold. The submanifold (a product of $M^2$ with an infinitesimal disc $D^2_R$)
     has been deleted.}
              }\label{sixd}} \;
\parbox{.45\textwidth}{
     \includegraphics[width=.49\textwidth]{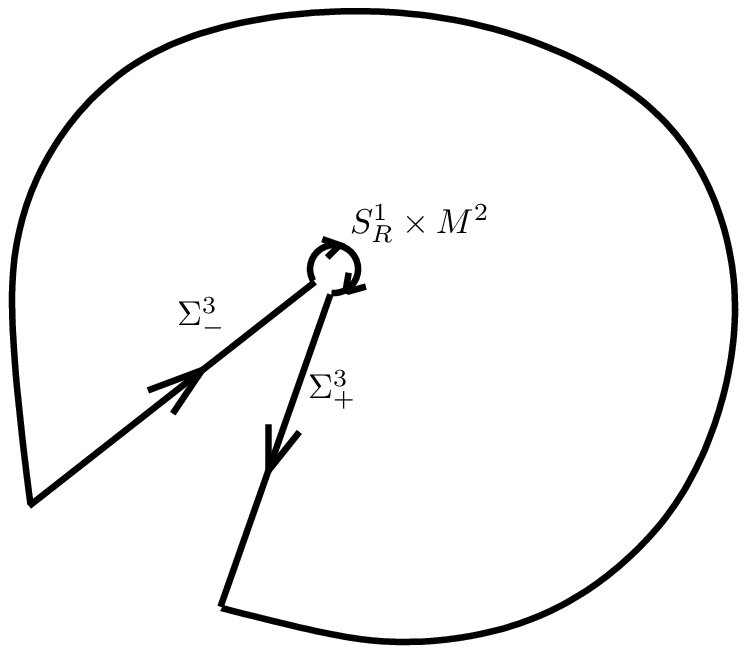}
     \caption{ {\small The geometry of fig \ref{sixd} is obtained by identifying $\Sigma^3_-$ with $\Sigma^3_+$
   and shrinking the radius, R, of the loop $S^1_R$ to zero.}
             }\label{cut}
    }
\end{figure}


\subsection{The two-dimensional case}


In order to describe the process in a simpler setting, let us consider the
case of a four-dimensional manifold $M^{4}$ with a two-dimensional defect,
as depicted in figures 1 and 2. For simplicity we will define $M^{4}$ as a
simply connected, non-compact, boundaryless manifold, such that it can be
covered by one chart. For example $M^4$ may have the topology of $\mathbb{R}%
^4$.

The action is given by the integral of the Characteristic form over $M^4 -M^2
$. We shall assume that $M^2$ is a two-dimensional submanifold without
boundary and furthermore we assume that it lies entirely in some
three-dimensional hyper-plane in $M^4$ (we will not consider the possibility
that the embedding of $M^2$ forms a non-trivial knot). The integral is
defined through the following regularisation process: From $M^4$ a tubular
neighbourhood $D_{R}^{2}\times M^{2}$ is removed, where $D_{R}^{2}$ is a
two-disk of radius $R$ with respect to some topological metric. We define a
four-dimensional integral over $M^{4}-M^{2}$ as the integral over $%
M^{4}-D_{R}^{2}\times M^{2},$ in the limit in which the radius $R,$ of the
2-disk $D_{R}^{2},$ goes to zero:
\begin{equation}
\int_{M^{4}-M^{2}}\left\langle \mathcal{FF}\right\rangle
:=\lim_{R\rightarrow 0}\int_{M^{4}-D_{R}^{2}\times M^{2}}\left\langle
\mathcal{FF}\right\rangle .  \label{lim2}
\end{equation}%
The excision of $D_{R}^{2}\times M^{2}$ from $M^{4}$ introduces the boundary
$\partial \left( M^{4}-D_{R}^{2}\times M^{2}\right) =S_{R}^{1}$ $\times
M^{2} $.

\begin{figure}[tb]
\begin{center}
\includegraphics[width=.9\textwidth]{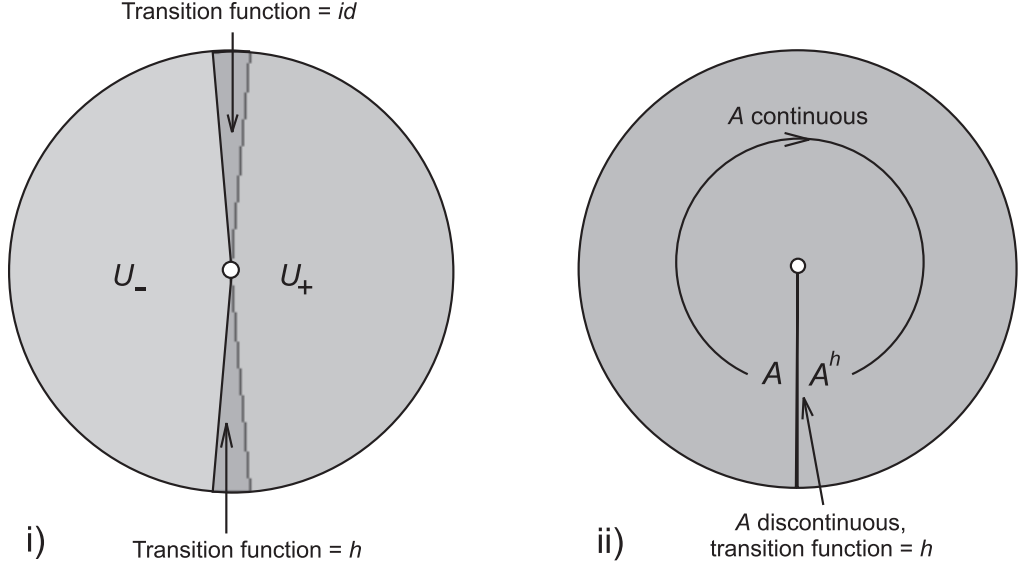}\\[0pt]
\end{center}
\caption{{\protect\small The manifold $M^{4}-M^{2}$ is covered by two charts
$U_{+}$ and $U_{-}$. They overlap in two disconnected regions as shown in
i). However, the transition function can be chosen to be trivial in one of
the regions. The other region can be shrunk to an effectively three
dimensional surface across which $A$ becomes discontinuous, as shown in ii).
}}
\label{twocharts}
\end{figure}

In view of the above assumptions about the topology of $M^4$ and
$M^2$, the domain of integration $M^{4}-D_{R}^{2}\times M^{2}$ can
be covered by two charts which are denoted by $U_+$ and $U_-$. The
overlap region, shown in figure \ref{twocharts} i), can be shrunk
to a three dimensional hyperplane which intersects the defect all
along the length of $M^2$. This hyperplane is divided into two
disconnected parts by the defect. The connections in each chart,
$A_+$ and $A_-$ respectively, are related by a transition function
in the overlap regions. However, since non-trivial holonomies can
only occur for paths which wind completely around the defect, it
is natural and convenient to take the transition function in one
of the overlap regions to be the identity. The other overlap
region is denoted $\Sigma$ and the
transition function is denoted $h$ . This is illustrated in figure \ref%
{twocharts} ii).
Thus, with this choice of atlas, the connection is continuous as one goes
around the defect except at $\Sigma $ where $\mathcal{A}_+$ and $\mathcal{A}%
_-$ are related by a gauge transformation:  $\mathcal{A}_+|_{\Sigma} =
\mathcal{A}_-^h|_{\Sigma}$,
\begin{equation}
\mathcal{A}^{h}:=h^{-1}\mathcal{A}h+h^{-1}dh\, .
\end{equation}

In each chart the Characteristic form can be expressed as a total derivative
so that $\int_{U_\pm} \langle \mathcal{F}\mathcal{F}\rangle = \int_{\partial
U_{\pm}} CS(\mathcal{A}_{\pm})$, where $CS(\mathcal{A})$ is the Chern-Simons
three form,
\begin{equation}
CS(\mathcal{A}):=\left\langle \mathcal{A}d\mathcal{A}+\frac{2}{3}\mathcal{A}%
^{3}\right\rangle\ .  \label{left}
\end{equation}%
The integral of (\ref{lim2}) thus reduces to an integral over the
contour depicted in figure 2,
\begin{eqnarray}
\lim_{R\rightarrow 0}\int_{M^{4}-D^{2}\times M^{2}}\left\langle \mathcal{FF}%
\right\rangle &=&\int_{\Sigma _{+}^{3}}CS(\mathcal{A}_-^{h})+\int_{\Sigma
_{-}^{3}}CS(\mathcal{A}_-)  \notag \\
&&+\lim_{R\rightarrow 0}\left\{ \int_{C^{+}\times M^{2}}CS(\mathcal{A}%
_+)+\int_{C^{-}\times M^{2}}CS(\mathcal{A}_-)\right\}  \label{LT}
\end{eqnarray}%
where $\partial U_{+}=\Sigma _{+}\bigcup (C^{+}\times M^2)$ and $\partial
U_{-}=\Sigma _{-}\bigcup (C^{-}\times M^2)$ and $C^\pm$ are semi-circles
such that $S_{R}^{1}=C^{+}\bigcup C^{-}$.

The first two integrals on the RHS of (\ref{LT}) correspond to the boundary
of the charts on the intersecting region. Defining the orientation of $%
\Sigma $ by $\Sigma \equiv -\Sigma _{-}\equiv \Sigma _{+}$ (and dropping the
subscript \textquotedblleft $-$" from $\mathcal{A}_{-}$), they become
\begin{equation}
\int_{\Sigma _{+}^{3}}CS(\mathcal{A}^{h})+\int_{\Sigma _{-}^{3}}CS(\mathcal{A%
})=\int_{\Sigma }CS(\mathcal{A}^{h})-CS(\mathcal{A})\,.  \label{C-C}
\end{equation}%
Now let us turn to the last two terms in (\ref{LT}). These two integrals
arise as further boundary terms along the $\mathbf{S}^{1}.$ The limit $%
R\rightarrow 0$ for these integrals seems to be, from a strict mathematical
point of view, somewhat ambiguous. Let us introduce a regularisation process
which will ensure that \emph{the integral on the RHS of (\ref{lim2}) is
invariant under gauge transformations $\mathcal{A}\rightarrow \mathcal{A}^{g}
$, for any $g(x)$ that is single valued in the limit that $S_{R}^{1}$
shrinks to a point.} We demand this because the integrand $\langle \mathcal{F%
}\mathcal{F}\rangle $ is gauge invariant and so the LHS should be invariant
under any such gauge transformation. This will be achieved if
\begin{equation}
\int\limits_{C^{+}\times M^{2}}CS(\mathcal{A}_{+})+\int\limits_{C^{-}\times
M^{2}}CS(\mathcal{A}_{-})\ \rightarrow \ \int\limits_{M^{2}}\langle \mathcal{%
A}\mathcal{A}^{h}\rangle \,.  \label{Claim}
\end{equation}%
As mentioned, the justification is ultimately the gauge invariance
of the final result. However, it is possible to obtain equation
(\ref{Claim}) by an adequate regularization, which is given in the
appendix.

Finally, setting that in the limit $\partial \Sigma =M^{2},$ and
using the identity $CS(\mathcal{A}^{h})\equiv
CS(\mathcal{A})-\frac{1}{3}\left\langle
\left( h^{-1}dh\right) ^{3}\right\rangle +d\left\langle h^{-1}\mathcal{A}%
dh\right\rangle $, allows writing (\ref{LT}) in a manifestly two-dimensional
form as
\begin{equation}
\int_{M^{4}-M^{2}}\left\langle \mathcal{FF}\right\rangle =-\int_{\Sigma }%
\frac{1}{3}\left\langle \left( h^{-1}dh\right) ^{3}\right\rangle
+\int_{M^{2}}\left\langle \left( \mathcal{A}-h^{-1}dh\right) \mathcal{A}%
^{h}\right\rangle .  \label{reduc}
\end{equation}%
The RHS of (\ref{reduc}) is a gWZW term, a two-dimensional action which has
the desired property of invariance under the local transformations,
\begin{equation}
h\rightarrow g^{-1}hg,\qquad \mathcal{A}\rightarrow g^{-1}\mathcal{A}%
g+g^{-1}dg,  \label{right}
\end{equation}%
that defines a theory on the topological defect, $M^{2}$. The field
equations, obtained by Euler-Lagrange variation with respect to $\mathcal{A}$
and $h$, are also invariant under the above gauge transformations.

It has been recognized that CS theory on a Riemann surface times $S^{1}$ is
equivalent to a WZW model \cite{Moore:1989yh}. The equality (\ref{reduc})
was conjectured to exist in \cite{Baez:1997jh}. We conclude that the two
dimensional action to be considered is\footnote{%
As usual, the action is defined up to a multiplicative constant.}
\begin{equation}
S(h,\mathcal{A})=\kappa \int_{\Sigma }\frac{1}{3}\left\langle \left(
h^{-1}dh\right) ^{3}\right\rangle -\kappa \int_{M^{2}}\left\langle \left(
\mathcal{A}-h^{-1}dh\right) \mathcal{A}^{h}\right\rangle ,
\end{equation}

The construction presented here generated a well known structure in two
dimensions starting from a four-dimensional topological invariant: the gWZW
terms that are the minimal gauge invariant extension of $\left\langle \left(
h^{-1}dh\right) ^{3}\right\rangle $. When a kinetic term for the Goldstone
fields is added, a good part of two dimensional physics can be retrieved
from this non-linear sigma model language: the description of the
super-string \cite{Henneaux:1984mh}; the characterization of exact string
backgrounds \cite{Witten:1991yr}; and the non-abelian bosonization phenomena
\cite{Witten:1983ar}, to name a few. The particular action described above,
the G/G model, is special in that, even when the kinetic term is added, it
defines a topological theory \cite{Witten:1991mm}. Thus, the $G/G$ model,
both with and without kinetic term, define very closely related theories, as
was discussed in Ref. \cite{Witten:1993xi}. In our construction, a kinetic
term does not arise.

This construction has produced a well defined action with all relative
coefficients fixed. The procedure can also be extended to build
gravitational actions in $2n-2$ dimensions beginning from the Euler density
in $2n$ dimensions.


\subsection{The four-dimensional case}


Applying the previous procedure to the six dimensional Euler density (\ref%
{Euler}) yields
\begin{equation}
S(h,\mathcal{A})=\frac{\kappa }{48\pi ^{3}}\int_{\Sigma }CS(\mathcal{A}) -CS(%
\mathcal{A}^{h}) - \frac{\kappa}{48\pi^{3}}\int_{M^{4}}B\left( \mathcal{A},
\mathcal{A}^{h}\right) ,  \label{1}
\end{equation}
where $CS(\mathcal{A})$ is now the CS five-form,
\begin{equation}
CS(\mathcal{A}):=\left\langle \mathcal{A}d\mathcal{A}d\mathcal{A}+\frac{3}{2}
\mathcal{A}^3 d\mathcal{A} + \frac{3}{5}\mathcal{A}^{5}\right\rangle ,\;
h=e^{\phi }=\exp (\frac{1}{2}J_{AB}\phi ^{AB}),
\end{equation}
and
\begin{equation}
B(\mathcal{A},\mathcal{A}^h):=\left\langle \mathcal{AA}^h\left( \mathcal{F} +%
\mathcal{F}^h- \frac{1}{2}\mathcal{A}^2- \frac{1}{2}\left(\mathcal{A}^h
\right)^2 +\frac{1}{2}\mathcal{AA}^h\right) \right\rangle .
\end{equation}
Replacing the identity,
\begin{eqnarray}
CS(\mathcal{A}) &\equiv &CS(\mathcal{A}^{h})+d\left\langle \left(
h^{-1}dh\right) \left(\mathcal{A}^h\mathcal{F}^h -\frac{1}{2}\left( \mathcal{%
A}^h\right)^3 \right)\right\rangle -\frac{1}{10}\left\langle\left(h^{-1}dh%
\right)^5 \right\rangle  \notag \\
&&-d\frac{1}{2}\left\langle \left( h^{-1}dh\right)^2 \mathcal{F}^h-
\left(h^{-1}dh\right) \mathcal{A}^h \left(h^{-1}dh\right) \mathcal{A}^h
\right\rangle  \notag \\
&&-d\frac{1}{2}\left\langle \left( h^{-1}dh\right) ^{3}\mathcal{A}%
^h\right\rangle
\end{eqnarray}
back in (\ref{1}) the action takes the form:
\begin{eqnarray}
S(h,A) &=&-\kappa \int_{\Sigma }\frac{1}{480\pi ^{3}}\left\langle
\left(h^{-1}dh\right) \left( h^{-1}dh\right) ^{2}\left(
h^{-1}dh\right)^2\right\rangle  \notag \\
&&+\frac{\kappa }{48\pi^3}\int_{M^4}\left\langle \left(dhh^{-1}\right)%
\mathcal{A}\left(d\mathcal{A}+ \frac{1}{2}\mathcal{A}^2\right) \right\rangle
\notag \\
&&-\frac{\kappa }{96\pi ^{3}}\int_{M^{4}}\left\langle \left( dhh^{-1}\right)
\mathcal{A}\left( \left( dhh^{-1}\right) ^{2} +\mathcal{A}%
\left(dhh^{-1}\right) \right) \right\rangle  \notag \\
&&-\frac{\kappa }{48\pi ^{3}}\int_{M^{4}}\left\langle \mathcal{AA}^{h}\left(%
\mathcal{F}+\mathcal{F}^{h}-\frac{1}{2}\mathcal{A}^{2}-\frac{1}{2}\left(%
\mathcal{A}^{h}\right) ^{2}+\frac{1}{4}[\mathcal{A},\mathcal{A}^{h}] \right)
\right\rangle  \label{2}
\end{eqnarray}

It must be stressed that the right normalization of the Wess-Zumino term was
obtained from the normalized Euler characteristic (\ref{Euler}) as a
by-product of the construction, without a need for adjusting the parameters
in the action (\ref{2}). The normalized Wess-Zumino term for a group with $%
\pi_5(G)=\mathbb{Z}$ satisfies \cite{Bott:1978bw}.
\begin{equation}
\int_{S^{5}}\frac{1}{480\pi^3}\left\langle \left(h^{-1}dh\right)^5
\right\rangle =n\in \mathbb{Z},  \label{n}
\end{equation}
where $n$ is the homotopy class to which the map $h:S^{5}\rightarrow G$
belongs.

Actions of the type (\ref{2}) are widely used in particle physics to
describe the infrared behavior of QCD \cite{Coleman:1969sm,Callan:1969sn}.
The gauged version was introduced originally by Witten in ref \cite%
{Witten:1988hc}, where the motivation was to find a gauge invariant
extension of the global $G\times G$ symmetry present in the five-dimensional
closed form $\left\langle \left( h^{-1}dh\right)^5\right\rangle$. This
problem is far from trivial, since the naive gauge extension of this term
obtained by replacing the exterior derivative by a covariant derivative
doesn't work: if this is done, the 5-form is no longer closed and the field
equations have support on the five-dimensional manifold $\Sigma $. Although
far from obvious, the same gWZW structures that arise in the description of
QCD may also be used to describe GR. While in QCD the gWZW term describes
the interactions of the infrared sector of the theory, here it might
correspond to an ultraviolet extension of GR.

The action (\ref{2}) was proposed as a gravitational model in \cite%
{Anabalon:2006fj} where, in order to obtain Einstein's field equations, a
field was fixed in the action. That is a rather unsatisfactory situation
since this is a condition imposed on a theory by an \textit{a posteriori}
expected result. In the next section we shall see that Einstein's field
equations arise from the action (\ref{2}) without fixing fields in the
action, but considering instead an ansatz that relies on the topological
defect interpretation of the action.

The field equations associated with the variation with respect to $h$ are
\begin{gather}
\int_{M^{4}}\Big\langle h^{-1}\delta h\Big\{\!\left( \mathcal{F}^h\right)^2 +%
\mathcal{F}^2+{}\!\mathcal{F}^{h}\mathcal{F}-\frac{3}{4} [{}\! \mathcal{A}%
^{h} -\mathcal{A},{}\mathcal{A}^{h}-\mathcal{A}]\ ({}\!\mathcal{F}^h +%
\mathcal{F})  \notag \\
+\frac{1}{8}[{}\!\mathcal{A}^{h}-\mathcal{A},{}\!\mathcal{A}^{h}-\mathcal{A}%
]^2 +\frac{1}{2}({}\!\mathcal{A}^{h}-\mathcal{A})[{}\!\mathcal{F}^{h} +
\mathcal{F},{}\!\mathcal{A}^{h}-\mathcal{A}])\Big\}\Big\rangle=0,
\label{EOM-h}
\end{gather}
while those associated with the connection $\mathcal{A}$ are
\begin{eqnarray}
0 &=&\int_{M^{4}}\Big\langle \delta \mathcal{A}\left( ({}\!\mathcal{A}^h -
\mathcal{A})\left( {}\!\mathcal{F}^h +2\mathcal{F} -\frac{1}{4}[{}\!\mathcal{%
A}^{h} -\mathcal{A},{}\!\mathcal{A}^h -\mathcal{A}]\right) \right) %
\Big\rangle  \notag \\
&&-(h\leftrightarrow h^{-1}).  \label{EOMA}
\end{eqnarray}

If one wishes to describe a four-dimensional world with Lorentzian
signature, the gauge group to be chosen can only be $SO(5,1)$, $SO(4,2)$ or $%
SO(3,3)$. The discussion will be restricted from now to the $SO(4,2)$ group,
as it is particularly interesting, allowing for the quantization of the
coefficient $\kappa$ in front of the action \cite{Zanelli:1994ti}. In the
next section we make contact between the action presented above and the
Einstein equations.


\section{The Einstein dynamical sector}


The topological action\footnote{%
Topological in the sense that no metric is needed to construct it.} (\ref{2}%
) gives rise to first order field equations, is invariant by construction
under coordinate transformations, and is also invariant under the local
transformations,
\begin{equation}
h\rightarrow g^{-1}hg,\;\;\mathcal{A}\rightarrow g^{-1}(\mathcal{A}+d)g.
\end{equation}
The theory contains 30 fields, the 15 components of $h\in SO(4,2)$, and 15
fields in the connection $\mathcal{A}=\frac{1}{2}\mathcal{A}^{AB}J_{AB}$.
The introduction of a four-dimensional topological defect in the
six-dimensional manifold splits the generators $J_{AB}$ into those that
leave invariant the tangent space of $M^{4}$, $J_{ab},J_{45}$, and those
that move it into the $4$ and $5$ directions, $J_{a4},J_{a5}$, where $%
a,b=0,...,3$ are Lorentz indices. It is therefore natural to separate the
generators into their irreducible Lorentz covariant parts $%
(J_{ab},J_{a5},J_{a4},J_{45})$. Correspondingly, the connection is written
as
\begin{equation}
\mathcal{A}=\frac{1}{2}\omega ^{ab}J_{ab}+c^{a}J_{a5}+b^{a}J_{a4}+\Phi
J_{45}\;,
\end{equation}
and the curvature reads
\begin{eqnarray}
\mathcal{F} &=&\frac{1}{2}(R^{ab}+c^{a}c^{b}-b^{a}b^{b})J_{ab}+[Db^{a}+c^{a}
\Phi ]J_{a4}+[Dc^{a}+b^{a}\Phi ]J_{a5}  \notag \\
&&+[d\Phi -b_{a}c^{a}]J_{45}.
\end{eqnarray}
Here $(J_{ab},J_{a5})$ and $(J_{ab},J_{a4})$ span the $so(3,2)$ and $so(4,1)$
subalgebras of $SO(4,2)$, respectively; $R^{ab}=d\omega
^{ab}+\omega_{\;c}^{a}\omega ^{cb}$ is the Lorentz curvature two-form and $%
Dc^{a}=dc^{a}+\omega _{\;b}^{a}c^{b}$. Note that the vielbein should be
identified as a vector under local Lorentz rotations. At this point there is
no strong reason to choose either $b$ or $c$, or any linear combination
thereof, as the vielbein.

In order to write down the field equations it is necessary to give a
parametrization of the group element. A convenient one can be constructed as
follows: take the Cartan decomposition $\mathfrak{g}=\mathfrak{p}\oplus
\mathfrak{q}$, where $\mathfrak{q}$ is the maximal compact subalgebra of $%
\mathfrak{g}$, $\mathfrak{p}=\mathfrak{g}-\mathfrak{q}$ and the semidirect
sum stands for $\left[ \mathfrak{p},\mathfrak{p}\right] \subset \mathfrak{q}$%
, $\left[ \mathfrak{q}, \mathfrak{p}\right] \subset \mathfrak{p}$, $\left[%
\mathfrak{q},\mathfrak{q}\right] \subset \mathfrak{q}$, the $so(4)$ indexes
are denoted by $\overline{a}=\left\{ 1,2,3,4\right\} $, so that $\mathfrak{q}
$ is spanned by $\left\{ J_{\overline{ab}},J_{05}\right\} $ and $\mathfrak{p}
$ by $\left\{ J_{5\overline{a}},J_{0\overline{a}}\right\} $, now due to this
decomposition any group element $g\in G$ can be written as $g=pq$, where $q$
is in the maximal compact proper subgroup of $G$, $q\in Q=SO(4)\times
SO(2)\subset G$, and $p$ is in its complement, $p\in P\subset G$. Any group
element of $P$ belongs to an orbit of the adjoint action of $Q$ on the
exponential of a Cartan subalgebra $\mathfrak{a}\subset \mathfrak{p}$ (see
for instance, \cite{Hermann}). Thus we have the decomposition of $G=QAQ$.
Applying this decomposition to $SU(2)$, for instance, gives the standard
parametrization in terms of the Euler angles and can be used in general to
decompose a given group in one parameter subgroups simplifying, in this way,
the computations. In our case, is enough to implement a partial
decomposition and the result is (See the appendix for more details)
\begin{equation}
h=h_{o}e^{\beta J_{12}}e^{\lambda J_{24}}e^{\delta J_{23}}e^{\left( z^{\bar{a%
}}J_{0\bar{a}}+\rho J_{52}\right) }e^{\alpha J_{05}}e^{\zeta ^{\overline{ab}%
}J_{\overline{ab}}}  \label{h}
\end{equation}
where $h_{o}$ is a constant group element whose effect corresponds to a
change in the origin of the parametrization. In our case, $h_{o}$
corresponds to the nontrivial identification that is made in the
six-dimensional manifold that gives rise to the defect. The presence of $%
h_{o}$ reflects the fact that the defect generates a non-dynamical
transition function of the six dimensional bundle. The fields $\beta,\lambda
,\delta ,z,\rho ,\alpha ,\zeta $, on the other hand, are fluctuations around
$h_{o}$. Since the directions transverse to the tangent space of the
topological defect are $4$ and $5$, the \textquotedblleft vacuum" of the
theory can be identified with the constant transition function $%
h_{o}=e^{\theta _{o}J_{45}}$.

On shell we fix $\lambda =\delta =\alpha =\beta =z=\rho =\zeta =0$. This
anzats simplifies the field equations enough to write them down by
components. From (\ref{EOMA}) it is straightforward to obtain (see appendix)
\begin{eqnarray}
\delta \Phi &:&0=0,  \label{a} \\
\delta c^a &:&\varepsilon_{abcd}c^b\left(3R^{cd}+ \left(2+\cosh
\theta_0\right) \left(c^c c^d -b^c b^d\right) \right) \sinh \theta_0=0,
\label{c} \\
\delta b^a &:&\varepsilon_{abcd}b^b \left(3R^{cd}+\left(2+\cosh
\theta_0\right) \left(c^c c^d -b^c b^d\right) \right) \sinh \theta_0=0,
\label{b} \\
\delta \omega^{ab} &:&3\varepsilon_{abcd}\left(b^c Db^d -c^c Dc^d\right)
\sinh \theta_0=0,  \label{d}
\end{eqnarray}
At this point it is clear that the choices of $b$ or $c$ as the vielbein
correspond to having a positive or negative cosmological constant,
respectively. In order to see that the Einstein equation are contained in
this system, it is sufficient to set $b=0$, keeping $c$ as the vielbein, and
requiring that $\theta _{0}\neq 0$. This further reduces the previous set of
equations to
\begin{eqnarray}
\varepsilon _{abcd}c^{b}\left( R^{cd}+\mu c^{c}c^{d}\right) &=&0,
\label{eins} \\
\varepsilon _{abcd}c^{c}Dc^{d} &=&0,  \label{tor}
\end{eqnarray}%
where $\mu =\frac{2+\cosh \theta_0}{3}$. Furthermore, the field equations
obtained varying with respect to $h$, (\ref{EOM-h}), are identically
satisfied by $\Phi =0$. This can be seen by substituting the ansatz (\ref{h}%
) into the field variations (\ref{EOM-h}). The components $(h^{-1}\delta
h)^{ab}$, $(h^{-1}\delta h)^{a4}$ and $(h^{-1}\delta h)^{a5}$ give field
equations proportional to the torsion $T^c = Dc^b$, and therefore are
identically satisfied by virtue of (\ref{tor}). The last component gives
\begin{align}  \label{Quadratic_Nasty}
(h^{-1} \delta h)^{45} \Big( R^{ab}+ \mu c^a c^b \Big) \Big( R^{cd}+ \mu c^c
c^d \Big)\varepsilon_{abcd} =0 \, .
\end{align}
Although this equation might seem to give a further restriction on the
geometry, that is not the case because $\left. (h^{-1} \delta h)^{45}
\right|_{h = h_0} = 0$, as can be easily verified for (\ref{h}). It must be
stressed, however, that this is not a property of the form chosen of the
parametrization (\ref{h}); any other parametrization obtained by gauge
transformation compatible with the presence of the defect would yield a
physically equivalent set of equations.

As in the three-dimensional case, when GR is regarded as a gauge theory \cite%
{Witten:1988hc}, contact with the metric phase of the theory makes it
necessary to require the vielbein to be invertible, $c_{\mu
}^{\;a}c_{\;a}^{\nu }=\delta _{\mu }^{\nu }$, $c_{\mu}^{\;a}c_{\;b}^{\mu
}=\delta _{b}^{a}$. The introduction of a parameter with dimensions of
length, $l$, is also necessary in order to make $\bar{c}_{\mu}^{a}=l^{-1}c_{%
\mu }^{a}$ dimensionless. These two conditions allow to regard $\bar{c}_{\mu
}^{a}$ as an isomorphism between the coordinate tangent space and the
non-coordinate one, such that the relation $g_{\mu \nu }=\bar{c}_{\mu }^{\;a}%
\bar{c}_{\nu }^{\;b}\eta_{ab}$ makes sense. Using this, equation (\ref{eins}%
) and the zero-torsion condition (\ref{tor}), reproduce the Einstein field
equations for the metric, $g_{\mu \nu }$,
\begin{equation}
R_{\mu \nu }-\frac{1}{2}g_{\mu \nu }R-\Lambda g_{\mu \nu}=0,
\end{equation}
where $R_{\mu \nu }$ is the metric-compatible Ricci tensor and $\Lambda
=l^{-2}\left( 2+\cosh \theta _{0}\right) $ is the cosmological constant.

However, the semiclasical description of gravitational solutions is also
related to the form of the action. In order to describe Einstein's gravity
as a minisuperspace of this gauged WZW theory, it is also necessary to
recover the Einstein Hilbert action. This can be done by replacing the
ansatz
\begin{equation}
\mathcal{A}=\frac{1}{2}\omega ^{ab}J_{ab}+c^{a}J_{a5}\;,\;\; h=e^{\theta
_{0}J_{45}}
\end{equation}
in the action (\ref{2}), reducing it to
\begin{equation}
\frac{\kappa \sinh \theta_0}{32\pi^3}\int_{M^4}\varepsilon_{abcd}c^{a}c^{b}%
\left( R^{cd}+\frac{1}{2}\mu c^{c}c^{d}\right) \, .
\end{equation}
This is indeed the Einstein-Hilbert action that gives GR with the same
cosmological constant that one obtains by putting the ansatz into the full
set of field equations, thereby justifying the use of a mini-superspace
action.


\section{Quantization of $\protect\kappa$ and Euclidean continuation}

\label{Euclidean_section}
The conditions under which the constant $\kappa $ takes integer values are
well known \cite{Witten:1983tw,Zanelli:1994ti}. Consider a non-linear sigma
model defined by the map
\begin{equation*}
h:S^{4}\longrightarrow G.
\end{equation*}
If $S^4$ is viewed as the boundary of some compact manifold $D^5$, one can
consider the extension of the map $h$ to the interior ($\partial D^{5}=S^{4}$%
),
\begin{equation*}
h:D^{5}\longrightarrow G.
\end{equation*}
However, $S^{4}$ can be the boundary of many different interiors. Imposing
independence of the path integral under a change of given interior $D^5$ by
another, $\overline{D}^5$ requires
\begin{equation}
S=\kappa \int_{D^5}\frac{1}{480\pi^3}\left\langle \left(h^{-1}dh\right)^5
\right\rangle= \kappa \int_{\overline{D}^5} \frac{1}{480\pi^3}\left\langle
\left(h^{-1}dh\right)^5 \right\rangle + 2\pi m \hbar.  \label{S}
\end{equation}
Therefore, the integral over the manifold, $M^5=D^5 \bigcup (-\overline{D}%
^5) $, where the minus denotes the correct orientation, is found to be
\begin{equation}
\kappa \int_{M^{5}}\frac{1}{480\pi ^{3}}\left\langle
\left(h^{-1}dh\right)^{5}\right\rangle= 2\pi m \hbar,
\end{equation}
where the manifold $M^5$ is compact and without boundary. Now if $\pi_{5}(G)=%
\mathbb{Z}$ then, from (\ref{n}), one concludes that $\kappa n = 2\pi
m\hslash$ and, as this must be true for all $n$, one concludes that $\kappa$
itself must be quantized,
\begin{equation}  \label{Qkappa}
\kappa= 2\pi n\hslash\ .
\end{equation}

This quantization holds for compact groups $G$, but the groups we are
considering here are not necessarily compact. However, there are complex
extensions of them that are compact. The argument presented here still holds
if one allows for analytic continuations of the theory, defined by the map
of the connection
\begin{eqnarray*}
\mathcal{A}^{\bar{a}\bar{b}} &\rightarrow &\overline{\mathcal{A}}^{\bar{a}%
\bar{b}}=\mathcal{A}^{\bar{a}\bar{b}},\; \mathcal{A}^{\bar{a}0}\rightarrow
\overline{\mathcal{A}}^{\bar{a}0}=i\overline{\mathcal{A}}^{\bar{a}0} \\
\mathcal{A}^{\bar{a}5} &\rightarrow &\overline{\mathcal{A}}^{\bar{a}5}=i%
\mathcal{A}^{\bar{a}5},\; \mathcal{A}^{05}\rightarrow \overline{\mathcal{A}}%
^{05}=-\overline{\mathcal{A}}^{05}.
\end{eqnarray*}
The same map must be applied to the Goldstone fields. The resulting action
is invariant under $SO(6)$, as can be seen by the equivalent map of the
generators
\begin{eqnarray}
J_{\bar{a}0} &\rightarrow &\bar{J}_{\bar{a}0}=iJ_{\bar{a}0},\;\;J_{\bar{a}%
5}\rightarrow \bar{J}_{\bar{a}5}=iJ_{\bar{a}5}  \notag \\
J_{05} &\rightarrow &\bar{J}_{05}=-J_{05},\;\;J_{\bar{a}\bar{b}}\rightarrow
\bar{J}_{\bar{a}\bar{b}}=J_{\bar{a}\bar{b}}
\end{eqnarray}%
where the new indexes $\bar{a}, \bar{b}$ cover the range $1,...,4$, and the
new metric is Euclidean, $\delta_{\bar{a}\bar{b}}=\left(+,+,+,+\right) $.
Under these changes, the invariant tensor $\langle J_{AB} J_{CD}
J_{EF}\rangle$ reverses sign and so does the action (\ref{2}), $S\rightarrow
-S$.

On the other hand, the Euclidean continuations of the groups $SO(5,1)$ and $%
SO(3,3)$ instead, give rise to additional imaginary factors in the action,
\begin{align*}
SO(5,1)\rightarrow SO(6)\quad &\Rightarrow\quad e^{\frac{i}{\hslash }\kappa
S} \rightarrow e^{-\frac{\kappa }{\hslash }S}\, , \\
SO(3,3)\rightarrow SO(6)\quad &\Rightarrow\quad e^{\frac{i}{\hslash }\kappa
S}\rightarrow e^{\frac{\kappa }{\hslash }S}.
\end{align*}

We see that the group $SO(4,2)$ has the particular property that since its
Euclidean continuation changes the action by a sign and not by an imaginary
factor, i.e.
\begin{gather}
SO(4,2)\rightarrow SO(6)\quad \Rightarrow\quad e^{\frac{i}{\hslash }\kappa
S} \rightarrow e^{-\frac{i}{\hslash }\kappa S}\, ,
\end{gather}
it allows for the existence of the phase freedom (\ref{Qkappa}). Conversely,
requiring the coefficient in front of the action to be quantized, singles
out the gauge group to be $SO(4,2)$.


\section{Discussion and Outlook}


Here, a six-dimensional gauge theory that gives rise to four-dimensional GR
has been proposed. The starting action (\ref{2}) is metric-independent, and
all the fields have a geometrical interpretation. Besides the usual
connection $\mathcal{A}$, the transition function $h$ around the
four-dimensional defect embedded in six dimensions is also present. These
two objects ($\mathcal{A}, h$) are completely defined once a principal
bundle is given over $M^6$.

The theory generalizes GR since it contains a dynamical sector in which
Einstein's equations hold, presumably reproducing all the experimental tests
that are compatible with GR. The Einstein-Hilbert Lagrangian is obtained as
the topological dimensional reduction of the six-dimensional Euler density
by the presence of the four-dimensional topological defect. In this way, a
theory that contains other fields besides GR is obtained, something that
could be welcome in the current state of affairs, where several models have
been advanced to explain the dynamics of the galaxies, inflation, or dark
matter in the Universe, and other phenomena that cannot be explained using
only GR and standard matter fields.

The purely gravitational sector studied here has classically zero torsion,
but the full theory naturally includes torsion. The presence of propagating
torsion in a background configuration changes many of the known results in
GR, including those about the generic existence of singularities in
spacetime \cite{Torsion}.

\begin{figure}[t]
\begin{center}
\includegraphics[height=3in]{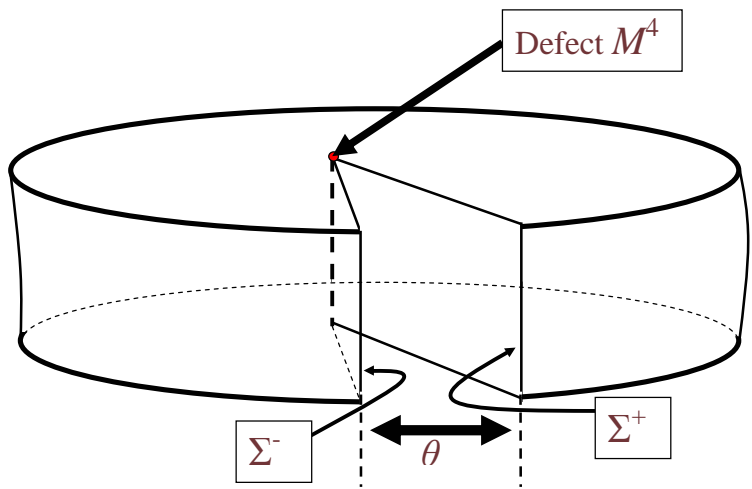}\\[0pt]
\end{center}
\caption{{\protect\small The physical interpretation of the transition
function $h = \exp(\protect\theta J_{45})$ as a defect caused by removing a
wedge from the six-dimensional manifold. It should be remembered that the
defect is parameterized by a hyperbolic angle, so this figure actually
represents the Euclideanized version discussed in section \protect\ref%
{Euclidean_section}. }}
\label{Topological_information}
\end{figure}
The transition functions represent topological information (fig.\ref%
{Topological_information}) of the six-dimensional action and become
dynamical in the four-dimensional theory. Their presence could be
interpreted as the deconfining phase of the higher dimensional, topological
theory and they could even be relevant to the description of our Universe.

The emergence of the space-time causal structure in the theory defined by (%
\ref{2}) arises only after a vielbein is chosen from amongst all the
invertible linear combinations of the $b$ and $c$.

Because of the non-trivial choice $h=\exp (\theta _{0}J_{45})$, the gauge
invariance of the theory is on shell reduced to SO(3,1)$\times$SO(1,1). The
choice $b^a =0$, $c^a \neq 0$ further breaks the SO(1,1) symmetry generated
by $J_{45}$, leaving the Lorentz group SO(3,1) as the remanent gauge
symmetry. The invertibility of what is chosen as a vielbein is not affected
by this remanent gauge symmetry: the vielbein $c^{a}$ transforms as a vector
under local Lorentz rotations.

The obtention of a gravitation theory that is metric independent; in which
GR could be seen as a broken phase of a topological field theory has been a
long sought goal \cite{Witten:1988xi}. The construction presented here is a
step in this direction. \newline

\textbf{Acknowledgments} \newline
The authors wish to thank Eloy Ay\'{o}n-Beato, Glenn Barnich, Fabrizio C\'{a}%
nfora, Steven Carlip, Frank Ferrari, Joaquim Gomis, Marc Henneaux, Matias
Leoni, Rafael Sorkin and Ricardo Troncoso for enlightening discussions.
Special thanks are given to Gast\'{o}n Giribet for his careful reading of
the manuscript, helping us to clarify several points. A.A. wishes to thank
the warm hospitality and the stimulating questions received at the General
Relativity workshop, Universidad de Buenos Aires (UBA) and the Instituto de
Astronom\'{\i}a y F\'{\i}sica del Espacio (IAFE), where this work was
presented; the support of MECESUP UCO 0209 and CONICYT grants is also
acknowledged. This work has been supported in part by FONDECYT grants Nos.
1040921, 1060831, 1061291, 3060016, and by an institutional grant to the
Centro de Estudios Cient\'{\i}ficos (CECS) from the Millennium Science
Initiative. Institutional support to CECS from Empresas CMPC is gratefully
acknowledged.

\appendix

\section{Appendix}

\setcounter{equation}{0} \numberwithin{equation}{section}

\paragraph{The regularisation process:\newline\newline}
 Here we give an argument to justify equation (\ref{Claim}). Let
$t$ be a coordinate on $S_{R}^{1}$ (anticlockwise) such that the
charts in $S^{1}$ are given by the range of coordinates
$C^{-}=(0,t_{\Sigma })$ and $C^{+}=(t_{\Sigma },1)$ where $t$ is
periodically identified $t\equiv t+1$. Let us introduce a family
of
functions $p_{n}(t)$ which, for each $n$, give a partition of unity on $%
S_{R}^{1}$ and which converge to the Heaviside step function $%
\lim_{n\rightarrow \infty }p_{n}(t)=\theta (t-t_{\Sigma })$. The
last two integrals in (\ref{LT}) can be expressed as:
\begin{equation}
\int\limits_{C^{+}\times
M^{2}}CS(\mathcal{A}_{+})+\int\limits_{C^{-}\times
M^{2}}CS(\mathcal{A}_{-})=\int\limits_{S_{R}^{1}\times
M^{2}}\left[ 1-p_{n}(t)\right]
CS(\mathcal{A}_{+})+p_{n}(t)\,CS(\mathcal{A}_{-})\,.
\end{equation}

Note that $\mathcal{A}_{\pm }$ in regions $C^{\pm }$ define a
non-trivial
bundle on $S_{R}^{1}$ that can not be extended to the interior of $D_{R}^{2}$%
. Let us define a new connection
$\mathcal{A}_{n}:=\mathcal{A}_{+}\left[ 1-p_{n}(t)\right]
+\mathcal{A}_{-}p_{n}(t)$ on $S_{R}^{1}$ which can be extended
into the interior since it is a single connection (continuous for
finite $n$, distributional for $n\rightarrow \infty $). Using the
property that $\lim_{n\rightarrow
\infty}p_{n}^{2}(t)=\lim_{n\rightarrow \infty}p_{n}^{3}(t)=\theta
(t-t_{\Sigma })$, it is possible to further show that:
\begin{align}
& \lim_{n\rightarrow \infty }\int\limits_{S_{R}^{1}\times
M^{2}}\left(
1-p_{n}\right) CS(\mathcal{A}_{+})+p_{n}\ CS(\mathcal{A}_{-})  \notag \\
& \qquad =\lim_{n\rightarrow \infty }\left[
\int\limits_{S_{R}^{1}\times
M^{2}}CS(\mathcal{A}_{n})-\int\limits_{S_{R}^{1}\times M^{2}}\dot{p}%
_{n}\langle {\mathcal{A}}_{+}\mathcal{A}_{-}\rangle \right]   \notag \\
& \qquad =\lim_{n\rightarrow \infty }\int\limits_{S_{R}^{1}\times M^{2}}CS(%
\mathcal{A}_{n})+\int\limits_{M^{2}}\left. \langle
\mathcal{AA}^{h}\rangle \right\vert _{t=t_{\Sigma }}\,.
\end{align}%
So for large $n$ we have
\begin{equation}
\int\limits_{C^{+}\times
M^{2}}CS(\mathcal{A}^{h})+\int\limits_{C^{-}\times
M^{2}}CS(\mathcal{A})\approx -\int\limits_{D_{R}^{2}\times
M^{2}}\langle
\mathcal{F}_{n}\mathcal{F}_{n}\rangle +\int\limits_{M^{2}}\langle \mathcal{AA%
}^{h}\rangle \, .
\end{equation}%
Now, being the manifold $D_{R}^{2}\times M^{2}$ regular and the curvature, $%
\mathcal{F}_{n}$, globally defined, it is reasonable to suppose
that the
integral of $\langle \mathcal{F}_{n}\mathcal{F}_{n}\rangle $ vanishes for $%
R\rightarrow 0$, in which case we recover equation (\ref{Claim}).

\bigskip

\paragraph{The following convention for the $SO(4,2)$ algebra was used:}

\begin{equation}
\left[ J_{AB},J_{CD}\right] =-J_{AC}\eta _{BD}+J_{BC}\eta _{AD}-J_{BD}\eta
_{AC}+J_{AD}\eta _{BC},
\end{equation}
\begin{equation}
A=0,...,5\ , \quad \eta _{AB}=\left( -,+,+,+,+,-\right) .
\end{equation}

\bigskip

\paragraph{Some notation: Einstein's Equation as a three-form.\newline
\newline
}

We have encountered the Einstein equation in differential form language. To
make contact with a more familiar tensorial form of the field equation, one
must require that the vielbein is not degenerate. The field equation is:
\begin{equation}
\varepsilon _{abcd}\ e^{b}\wedge R^{cd}\ =\ \frac{1}{2}\varepsilon _{abcd}\
R_{\ \ ef}^{cd}\ e^{b}\wedge e^{e}\wedge e^{f}=0\,.  \label{Z}
\end{equation}%
Then the trick is simply to dualise this three form:
\begin{equation*}
\varepsilon _{abcd}\ R_{\ \ ef}^{cd}\ \varepsilon ^{befg}=0
\end{equation*}%
Now one makes use of the identity $\varepsilon _{abcd}\ \varepsilon
^{befg}=\delta _{acd}^{efg}$, where the $\delta $ is the totally
antisymmetrised Kronecker delta. One obtains
\begin{equation*}
R_{\ \nu }^{\mu }-\frac{1}{2}\delta _{\ \nu }^{\mu }\ R=0,
\end{equation*}%
To avoid notational confusion ($R_{\ \nu }^{\mu }$ is the Ricci tensor, not
to be confused with $R^{ab}$ which is the curvature two-form) the indices
have been converted to coordinate basis indices using the vielbein $e_{\mu
}^{a}$ and its inverse. The familiar tensorial form of Einstein's field
equation is thus recovered.

\bigskip

\paragraph{Some useful formulas\newline
\newline
}

Given $h=e^{\theta J_{45}}$, it is possible to compute:
\begin{eqnarray}
\mathcal{A}^{h} &=&\frac{1}{2}\omega _{ab}J^{ab}+J_{a4}\left( b^{a}\cosh
\theta +c^{a}\sinh \theta \right) +J_{a5}\left( c^{a}\cosh \theta
+b^{a}\sinh \theta \right)  \notag \\
&&+\left( \Phi +d\theta \right) J_{45}
\end{eqnarray}

\begin{eqnarray*}
\mathcal{F}^{h} &=&\frac{1}{2}\left( R^{ab}+c^{a}c^{b}-b^{a}b^{b}\right)
J_{ab}+J_{a4}\left[ \left( Db^{a}+c^{a}\Phi \right) \cosh \theta +\sinh
\theta \left( Dc^{a}+b^{a}\Phi \right) \right] \\
&&+J_{a5}\left[ \left( Dc^{a}+b^{a}\Phi \right) \cosh \theta +\sinh \theta
\left( Db^{a}+c^{a}\Phi \right) \right] +\left( d\Phi -b^{a}c_{a}\right)
J_{45}
\end{eqnarray*}

\begin{align}
\left[ \mathcal{A}^{h}-\mathcal{A},\mathcal{A}^{h}-\mathcal{A}\right]
&=2\left( 1-\cosh \theta \right) \left( c^{a}c^{b}-b^{a}b^{b}\right) J_{ab}
\notag \\
& \quad +2\left( \left( \cosh \theta -1\right) c^{a}+\sinh \theta
b^{a}\right) d\theta J_{a4}  \notag \\
& \quad +2\left( \left( \cosh \theta -1\right) b^{a}+\sinh \theta
c^{a}\right) d\theta J_{a5}  \notag \\
& \quad +4\left( 1-\cosh \theta \right) c^{a}b^{b}\eta _{ab}\, J_{45}
\end{align}

\bigskip

\paragraph{On the group parametrization\newline
\newline
}

How the parametrization used in this paper arise from the Cartan
discomposition,

\begin{equation}
h=h_{o}e^{x^{\bar{a}}J_{5\bar{a}}+y^{\bar{a}}J_{0\bar{a}}}k\
\end{equation}%
where $k$ is an arbitrary group element of the maximal compact subgroup of $%
SO(4,2)$, can be explicitly checked as follows: first note that

\begin{equation}
e^{\beta J_{12}}e^{\lambda J_{24}}e^{\delta J_{23}}\rho J_{52}e^{-\delta
J_{23}}e^{-\lambda J_{24}}e^{-\beta J_{12}}=x^{\bar{a}}J_{5\bar{a}}
\end{equation}%
where
\begin{eqnarray}
x^{1} &=&\rho \sin \beta \cos \delta \cos \lambda, \;  \notag \\
x^{2}&=&\rho \cos \beta \cos \delta \cos \lambda, \;  \notag \\
x^{3} &=&-\rho \sin \delta \; x^{4}=-\rho \cos \delta \sin \lambda .
\end{eqnarray}%
It follows that

\begin{equation}
x^{\bar{a}}J_{5\bar{a}}+y^{\bar{a}}J_{0\bar{a}}=e^{\beta J_{12}}e^{\lambda
J_{24}}e^{\delta J_{23}}\left( z^{\bar{a}}J_{0\bar{a}}+\rho J_{52}\right)
e^{-\delta J_{23}}e^{-\lambda J_{24}}e^{-\beta J_{12}}
\end{equation}%
where the redefinition in the coordinates

\begin{eqnarray}
z^{1} &=&y^{1}\cos \beta -y^{2}\sin \beta  \notag \\
z^{2} &=&y^{1}\cos \lambda \sin \beta \cos \delta +y^{2}\cos \beta \cos
\lambda \cos \delta -y^{3}\sin \delta -y^{4}\cos \delta \sin \lambda  \notag
\\
z^{3} &=&y^{1}\sin \beta \cos \lambda \sin \delta +y^{2}\cos \beta \cos
\lambda \sin \delta +y^{3}\cos \delta -y^{4}\sin \lambda \sin \delta  \notag
\\
z^{4} &=&y^{1}\sin \beta \sin \lambda +y^{2}\cos \beta \sin \lambda
+y^{4}\cos \lambda
\end{eqnarray}%
was used.\ Finally the group element can be written as\newline

\begin{equation}
h=h_{o}e^{\beta J_{12}}e^{\lambda J_{24}}e^{\delta J_{23}}e^{\left( z^{\bar{a%
}}J_{0\bar{a}}+\rho J_{52}\right) }k_{1}\
\end{equation}%
where $k_{1}=e^{-\delta J_{23}}e^{-\lambda J_{24}}e^{-\beta J_{12}}k$ is an
arbitrary compact subgroup element.

\end{document}